\documentstyle[amssymb,twocolumn,prl,aps]{revtex}

\begin{document}

\twocolumn[
\hsize\textwidth\columnwidth\hsize\csname@twocolumnfalse\endcsname
\draft
\title{New Type of Charge and Magnetic order in the Ferromagnetic Kondo Lattice }
\author{D. J. Garc{\'\i}a$^{(1)}$, K. Hallberg$^{(1)}$, C. D. Batista$^{(1)}$, M. Avignon$^{(2)}$ and B. Alascio$^{(1)}$}
\address{(1)Centro At\'omico Bariloche and Instituto Balseiro}
\address{Comisi\'on Nacional de Energ{\'\i}a At\'omica, 8400 S.C. de Bariloche, Argentina}
\address{(2)Laboratoire d' Etudes des Propri\'et\'es Electroniques des Solides(LEPES)}
\address{Centre National de la Recherche Scientifique (CNRS), BP 166, 38042 Grenoble Cedex 9, France}
\date{Received \today }
\maketitle

\begin{abstract}
We study numerically the one dimensional ferromagnetic Kondo lattice, a model
widely used to describe nickel and manganese perovskites.  By including a
nearest neighbor Coulomb interaction (V) and a superexchange interaction
between the localized moments (K), we obtain the phase diagram in parameter
space for several dopings at T=0.  Due to the competition between double and
superexchange, we find a region where the formation of magnetic
polarons induces a charge ordered (CO) state which survives also for V=0. This
mechanism should be taken into account in theories of charge ordering
 involving spin degrees of freedom.
\end{abstract}

]

\narrowtext

In recent years there has been great interest in the nontrivial interplay of
charge, spin and lattice degrees of freedom in strongly correlated electron
systems, especially in perovskite transition-metal oxides. One of the most
striking phenomena is the simultaneous appearance of charge and spin
superstructures. For example, neutron scattering\cite{hayden} and electron
diffraction\cite{chen} experiments in La$_{2-x}$Sr$_{x}$NiO$_{4}$ showed the
presence of charge/lattice and spin modulations with doping-dependent wave
vector. Stripe formation together with incommensurate spin fluctuations in
High Tc superconductors can also be regarded as a manifestation of similar
phenomena\cite{tranq}. The charge and spin ordering found in many of the
doped manganese perovskites also fall in the same category. Experiments have
revealed CO at half filling in Nd$_{0.5}$Sr$_{0.5}$MnO$_{3}$\cite{kawano}
and similar compounds, such as Pr$_{0.5}$Ca$_{0.5}$MnO$_{3}$,\cite
{tomioka,li,moritomo,chen2}. More recent interest has focused on electron
doped charge ordered manganites \cite{liubao,mori,ueharajirac}. CO has also
been found for other dopings as in Bi$_{1-x}$Ca$_{x}$MnO$_{3}$\cite{liubao}
and in La$_{1-x}$Ca$_{x}$Mn$O_{3}$ (doped with $Pr$) \cite{ueharajirac} for $%
x\geqq 0.5$ (few electron region).

The ferromagnetic staircase structure of the CE phase is found in several of
the $x=0.5$ manganites and the CO found in these have been interpreted in
terms of orbital ordering \cite{Khomski,Dagotto}. The ordering between
chains, however, is not yet clear.

Although there have been several attempts to explain the CO phase
theoretically in the half-filled case by considering two $Mn$ orbitals\cite
{mizokawa,jackeli,Khomski} with and without intersite Coulomb interaction $V,
$ and adding strong on-site Coulomb interactions\cite{Khomski}, there
remains, to the best of our knowledge, no explanation for the existence of
CO in the electron-doped region ($x>0.5$) \cite{liubao,mori} . Numerical
studies for several dopings\cite{malvezzi} have included the effect of $V$
and obtained a very rich phase diagram, finding phase separated (for either
extreme dopings) and CO ($x\simeq 0.5$) regimes. In Ref. \cite{yunoki} the $%
x=0.5$ two-orbital case is also studied using Monte Carlo techniques, and
the CO phase is stabilized by Jahn-Teller phonons. As mentioned in this
work, the $z$-axis stacking of charge and existence of bistripes at $x>0.5$%
\cite{mori}, both penalized by a large Coulomb interaction, indicate that $V$
is smaller than expected and is not enough to understand the CO state.

Unlike CO in non-magnetic materials where the Peierls instability or large
intersite Coulomb interactions are required, we will show here that in these
magnetic materials charge density waves can result from the formation of
magnetic superstructures arising from the presence of competing
interactions. By changing the carrier concentration or the relationship
between the competing interactions, this new mechanism gives rise to a very
rich family of inhomogeneous spin and charge structures.

In order to illustrate this new mechanism we will study here a simplified
model where the competing forces are personified by the double and
superexchange interactions. The Ferromagnetic Kondo Lattice Model (FKLM),
which was devised for manganites, was studied originally by de Gennes\cite
{de gennes}. The most intriguing question is what happens in an intermediate
regime, where the competing interactions are energetically similar. De
Gennes proposed canting of two interpenetrating lattices as the compromise
solution of this competition. This concept was also used subsequently in
recent analytical approaches \cite{Aliaga,Arovas}. Phase separation has also
been considered as a possible solution to this competition \cite{Kagan,Moreo}%
.

As will be shown below, we find spin phases which cannot be described in
terms of two interpenetrating sublattices nor do they correspond to phase
separation.

As a result of the competition between the double exchange (DE) mechanism
which delocalizes the hole and the superexchange (SE) between local spins,
different phases may appear such as ferromagnetism in one extreme,
antiferromagnetism in the other, and doping-dependent modulated charge and
spin order in between. For $x=1/2,$ we find no charge ordering for $V=0,$
but away from that concentration we find charge modulation even in the
absence of Coulomb interactions. In this case, we find what can be described
as an ordered phase of ferromagnetic islands in which carriers are
localized. These phases are insulating, but a magnetic field can induce a
transition to a metallic phase.

We consider the widely used FKLM. For completeness we add the effect of
nearest-neighbors intersite interaction when considering $x=0.5$:

\begin{eqnarray}
H &=&-t\!\sum_{i}(c_{i\sigma }^{\dagger }c_{i+1\sigma
}+h.c.)+U\!\sum_{i}n_{i\uparrow }n_{i\downarrow }+  \nonumber \\
&&V\!\sum_{i}n_{i}n_{i+1}+J_{h}\!\sum_{i}{\bf S}_{i}{\bf \sigma }%
_{i}+K\!\sum_{i}{\bf S}_{i}{\bf S}_{i+1}
\end{eqnarray}

Here the first term represents the $e_{g}$-electron transfer between
nearest-neighbor Mn ions at sites $i$ and $i+1$ (we will take $t$ as the
unit of energy), the second and third terms are the on-site and inter-site
Coulomb repulsions between these orbitals, $J_{h}$ is the Hund's rule
coupling between localized ${\bf S}_{i},$ taken to have $S=1/2$, and
itinerant ${\bf \sigma }_{i}$ spins and $K$ is the superexchange between
local spins.

We will fix the values $J_{h}=20$ and $U=10$, so that the only free
parameters are $V$ and $K$. For the large $J_h$ used in this paper the results are not sensitive to $U$
since double occupancy is supressed by $J_h$. We consider here the case of one non-degenerated 
$e_{g}$ orbital, and defer the analysis of the role of orbital ordering for
future study. The ground state of this model has also been calculated by
other authors \cite{malvezzi} using the finite-system DMRG \cite{white,Karen}%
. Here we have taken special care in the growing procedure in order not to
frustrate the system in the cases when charge and spin order are expected 
\cite{Daniel}. With this consideration, very accurate results were obtained
with a discarded weight lower than $10^{-4}$ for the largest systems
presented here.

Let us consider first the case $V=0,$ in order to isolate the two competing
interactions $K$ and $t$. For any finite concentration there is a fully
polarized ferromagnetic (F) phase below some critical value of $K/t,$ and an
antiferromagnetic (AF) phase for sufficiently large $K/t$. This can be seen
in Fig. 1 for $x=0.5$. In the intermediate regime we get a phase ($%
0.2<K/t<0.4$ ) with clear peaks in the spin structure factor $S(q)$ at $%
q=\pm \pi /2$ (see Fig. 1a) . These peaks increase logarithmically with $L$
(see Fig. 2a) indicating a power law decay of the spin-spin correlation
function in real space. This power law decay is a consequence of the fact
that no long range order can be sustained in a one dimensional model with $%
SU(2)$ symmetry. This structure has been obtained with quantum Montecarlo
using classical spins and interpreted as a spiral state with pitch $q=\pi /2$%
\cite{yuno1}. Instead our results on the real space spin-spin correlation
function show ferromagnetic pairs coupled antiferromagnetically:($\uparrow
\uparrow \downarrow \downarrow \uparrow \uparrow \downarrow \downarrow ...$%
). Our classical Montecarlo calculations \cite{Horacio} confirm this picture.

Now if we look at the charge, we find that even though there is no charge
ordering in any of these phases, the charge-charge correlation functions
shown in Fig 1b are essentially different. While the ferromagnetic case ($%
K/t=0.1$) corresponds to a spinless metallic phase (the peak at $%
q=2k_{F}=\pi $ is a consequence of Friedel oscillations), in the AF case ($%
K/t=1$) $N(q=\pi )$ the Friedel oscillations loose weight (see Fig. 1).
Concerning the $q=\pi /2$ phase, it can be seen from Fig. 1b and Fig. 2b
that there is no charge ordering for this phase, even when the charge-charge
correlation function is quite different compared to the ferromagnetic case.
This change can be understood if we consider that the charges are
`localized' in bonds due to the spin structure ($\uparrow \uparrow
\downarrow \downarrow \uparrow \uparrow \downarrow \downarrow ...$), which
would give an insulating character to this phase (preliminar calculations of
the Drude weight seem to confirm this view \cite{Poilblanc}). Each charge
induces a ferromagnetic island of two spins and tends to be localized in the
bond in order to gain kinetic energy. The shape of $N(q)$ is closer to $%
(1-\cos q)/4$ as a consequence of the enhancement of the charge correlations
within an island. From this point of view it is easy to understand the
absence of charge ordering, because both sites are completely equivalent in
each island. However, this would not be the case if the tendency to form
ferromagnetic islands surrounding each charge were preserved for lower
electron concentrations. If we consider, for example, $x=2/3$ the expected
spin phase may be represented schematically as $\uparrow \uparrow \uparrow
\downarrow \downarrow \downarrow \uparrow \uparrow \uparrow \downarrow
\downarrow \downarrow ...$, where the spins cluster in the form of magnetic
polarons\cite{Nos1,Nos2}. In this case, the centers of each island are not
equivalent to the borders, and the charge will tend to accumulate in the
middle of the polarons. For this reason we expect to get CO together with a
spin density wave for electron concentrations lower than $0.5$. This scheme
is confirmed by Fig. 3, where we show the charge and spin structure factors
calculated for $x=2/3$ and $K/t=0.25$. $S(q)$ clearly shows a peak at $q=\pi
/3$, while $N(q)$ has a pronounced peak at $q=2\pi /3$ (the peak at small $q$ in $N(q)$ for the $x=1/3$ case is due to a kink in
the center of the open chain). Fig.2 shows that the
intensity of these peaks scale logarithmically with $L$ for $S(q=\pi /3)$
and linearly for $N(q=2\pi /3)$ which is a clear evidence of long-range
charge ordering. The same calculations for $x=0.2,0.25$ show spin and charge
correlation functions consistent with the formation of 4- and 5-sites spin
islands containing one electron each. These structures could be regarded as
a crystallization of the magnetic polarons described in references \cite
{Nos1,Nos2} for the dilute limit. It is interesting to note that CO is
induced by spin ordering and vice versa, demostrating that the formation of
these superstructures is a consequence of the interplay between charge and
spin degrees of freedom.

In the case $x=1/3,$ it is evident from Figs. 2 and 3 that there is also a
spin and charge-ordered state for this concentration. Both $N(q)$ and $S(q)$
now have peaks at $q=2\pi /3$. This result can be understood clearly with
the following image for the magnetic structure: $\uparrow \uparrow
\downarrow \uparrow \uparrow \downarrow \uparrow \uparrow \downarrow $,
while the charge is distributed with one electron in each pair of up spin
sites, and one in each down spin site.

The validity of the image described above can be easily tested using a
classical approach. We calculate the contributions of double exchange and
superexchange to the energy of the different possible phases for each value
of the concentration, assuming that the hopping vanishes at the
antiferromagnetic bonds. For example in the case {\bf \ }$x=1/2,$ we obtain $%
-2t/\pi +K$ for the ferromagnetic phase, $-t/2$ for the $q=\pi /2\ (\uparrow
\uparrow \downarrow \downarrow \uparrow \uparrow \downarrow \downarrow )$
phase, and $-K$ for the antiferromagnetic phase. The sequence of stable
phases when increasing $K/t$, F$\rightarrow \pi /2\rightarrow $AF is
obtained in agreement with the numerical results. In this classical picture
a canted AF phase has a slightly lower energy than the pure AF phase, but
the $\pi /2$ phase remains stable over a certain region in $K/t$. The same
procedure leads to similar conclusions at $x=1/3$ and $2/3.$

It is also of interest to consider the effect of an intersite Coulomb
interaction $V$ at $x=1/2$, because at this concentration it is most
effective in inducing charge ordering. All phases exhibit charge ordering
above a critical value of $V$. In Fig.4 we show a tentative phase diagram in 
$V$ and $K$. Our calculations for different values of $K$ indicate a
first-order transition from the fully polarized ferromagnetic phase to the
one with $q=\pi /2$ spin order. This metal-insulator transition can be
induced by the application of a magnetic field. As we mentioned above this insulator to metal transition can also be
obtained by applying a magnetic field that polarizes the spins, thus
delocalizing the electrons due to the double exchange interaction. At higher $K/t$ the latter
evolves into the AF phase.

The Coulomb repulsion $V$ inhibits the double-exchange mechanism by reducing
the mobility of the carriers. This reduces the phase space for the
ferromagnetic order. As $K$ increases, the critical value of $V$ for charge
ordering diminishes from the spinless value $2t$ valid for the saturated
ferromagnetic regime. For{\em \ }$V>>t${\em \ }we obtain the ferrimagnetic
CO (CO-FIM) phase{\em \ ...}$\Uparrow \downarrow \Uparrow \downarrow \Uparrow
\downarrow ...$of alternating $S=1/2$ and $S=1$ spins.

In summary, we have presented numerical evidence of the existence of a new
type of simultaneous charge and spin ordering in the FKLM. This mechanism,
induced by the competition between DE and SE, is based on a striking
interplay between charge and spin degrees of freedom. As we have shown in
previous works \cite{Nos1,Nos2}, in the dilute limit each carrier polarizes
its surroundings forming ferromagnetic polarons. The size of these polarons
is governed by the ratio $K/t$. We could interpret our results as an
indication that if this size is similar to the mean separation between
carriers, then SE tends to order these islands separating them by
antiferromagnetic interphases. Of course, this is an approximate image
because there are quantum magnetic fluctuations between and inside the islands
. While previous work reported the need of coulomb interaction $V$ \cite
{malvezzi} or electron-phonon coupling $\lambda $ \cite{yunokihotta} to
stabilize a charge-ordered phase, we show that it arises naturally
considering the interplay between charge and spin degrees of freedom.

A great variety of experiments have found the presence of charge and spin
ordering in manganites. The extraordinary colossal magnetoresistance effect
for La$_{0.5}$Ca$_{0.5}$MnO$_{3}$ involves the abrupt destabilization of a
CO-AF state by a magnetic field \cite{Tomioka}. Insulating charge-ordered
and metallic ferromagnetic regions coexist in (La$_{0.5}$Nd$_{0.5}$)$_{2/3}$%
Ca$_{1/3}$Mn$O_{3}$ \cite{Ibarra} and Pr$_{0.7}$Ca$_{0.3}$MnO$_{3}$ \cite
{Cox,Kiryukhin}. Both phenomena indicate that the CO phase is very close in
energy to the FM\ state. In addition, the z-axis stacking of charge and
existence of bistripes at $x>1/2$ \cite{mori}, indicate that CO is not
driven by a Coulomb repulsion. These observations are consistent with a
picture where spin and charge ordering are driven by the strong interplay
between charge and spin degrees of freedom. For these reasons, the
phenomenon presented here could be the underlying physical mechanism for the
stabilization of spin and charge structures in manganites.

Our model calculations fail to include several effects that may play
important roles in real systems like Jahn-Teller distortions and orbital
degeneracy. However it is important to understand clearly that charge
ordering can be induced simply as a result of the competition between DE and
SE as shown here. Considerable work remains to be done. For example, it
would be desirable to complete a $K/t$ vs. $n$ phase diagram.

We thank Silvia Bacci for helping us with computer programs and B. Normand
and D. Poilblanc for useful sugestions and discussions. Three of us (D. G.,
C.D.B. and K. H.) are supported by the Consejo Nacional de Investigaciones
Cient\'{i}ficas y T\'{e}cnicas (CONICET). B. A. is partially supported by
CONICET. We would like to acknowledge support from the Program for
scientific collaboration between France and Argentina ECOS-SECyT A97E05 and
the University of Buenos Aires where part of this work has been done.


Figure 1: Spin and charge structure factors for $x=0.5$ and $L=28$.

Figure 2: Size dependence of the peaks of the spin and charge structure
factors.

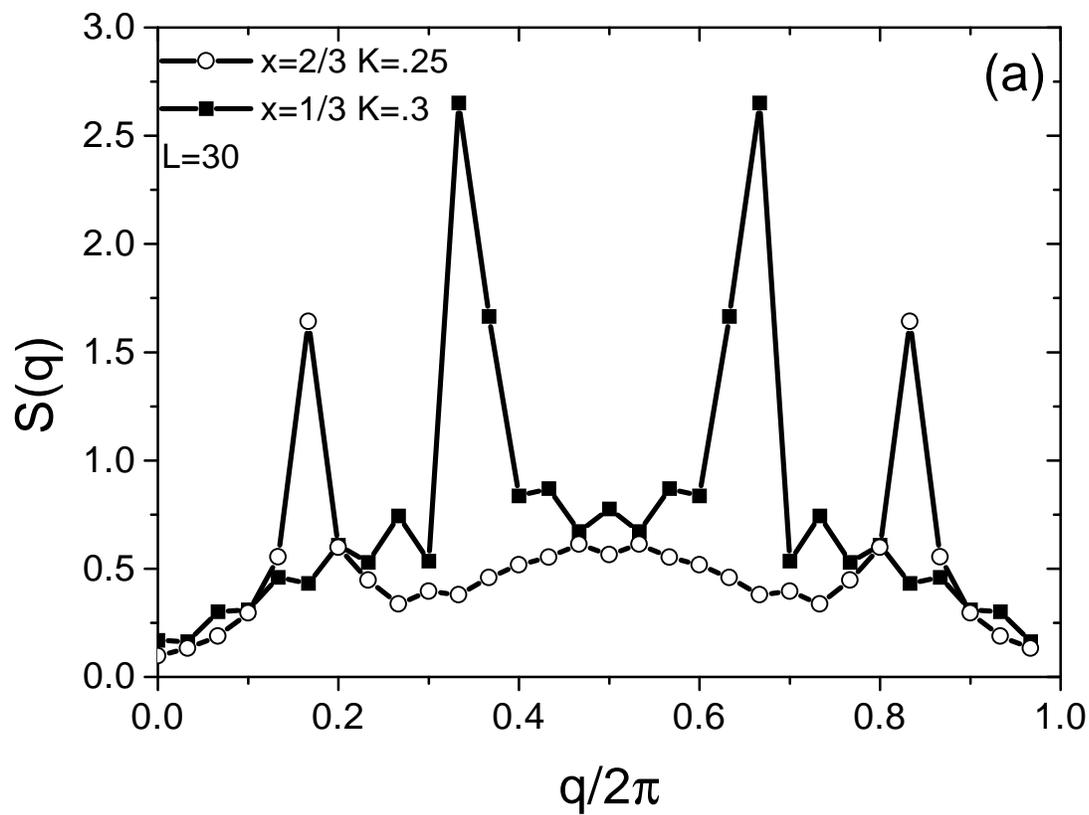
Figure 3: Spin and charge structure factorsfor $x=1/3,2/3$ and $L=30.$

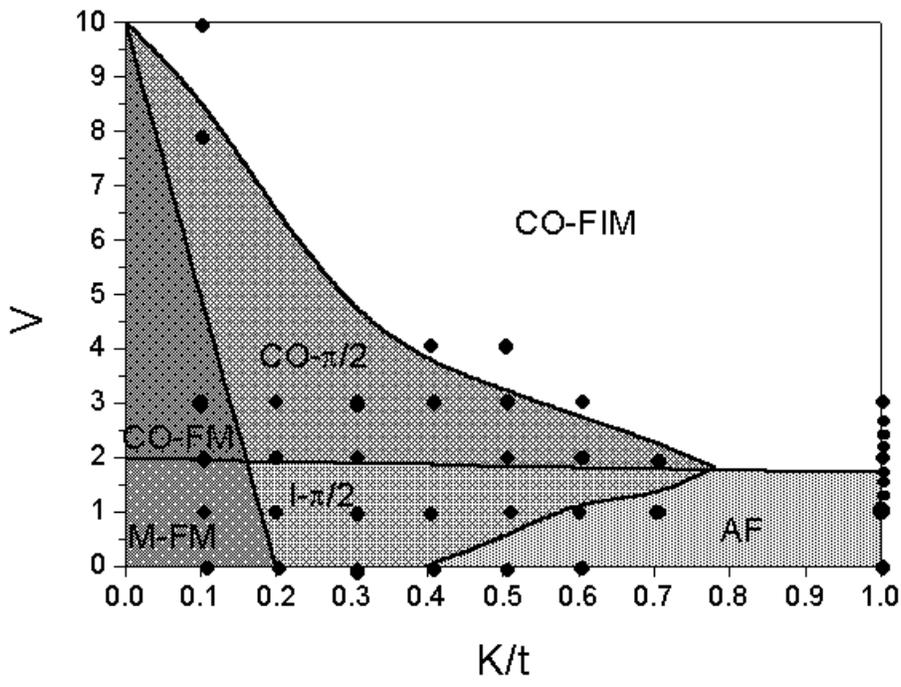
Figure 4: Phase diagram K/t vs. V/t for $x=0.5$. The circles indicate the
points where calculations have been performed.

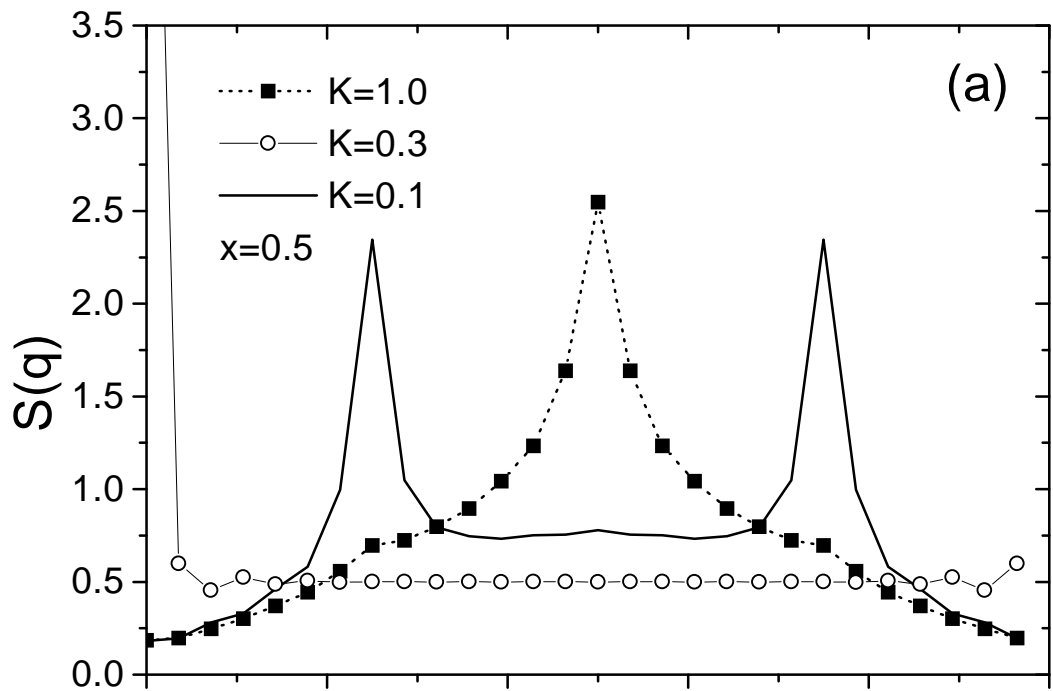
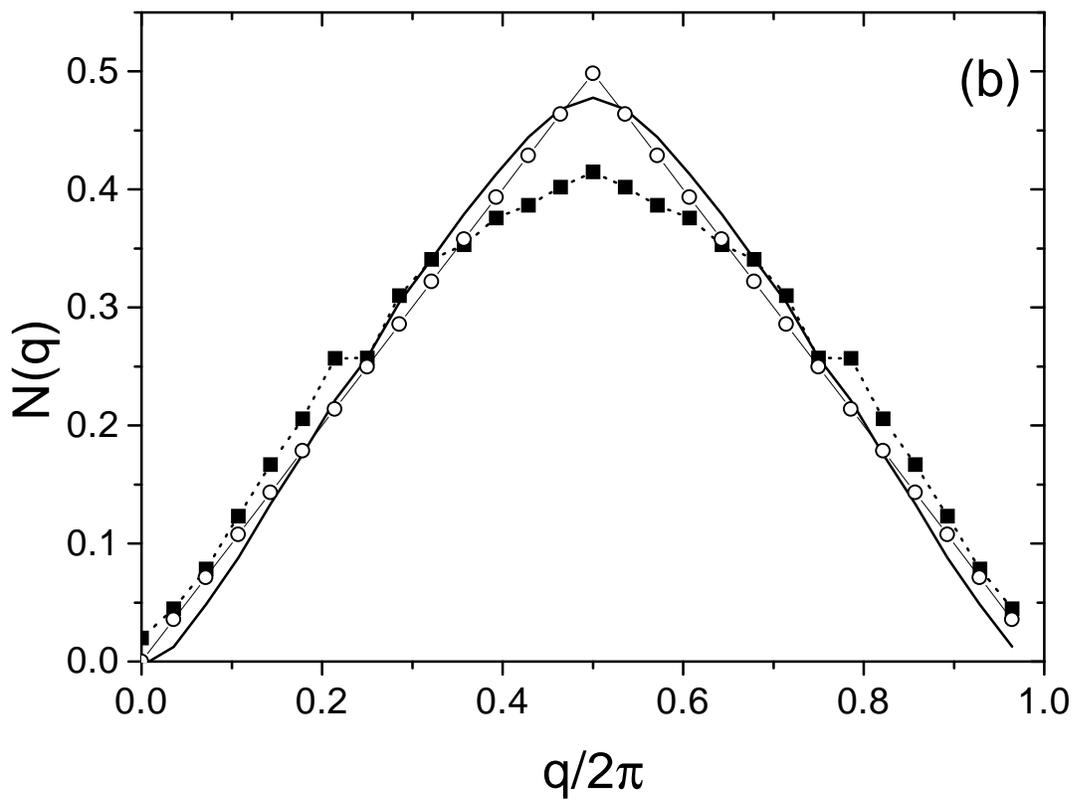
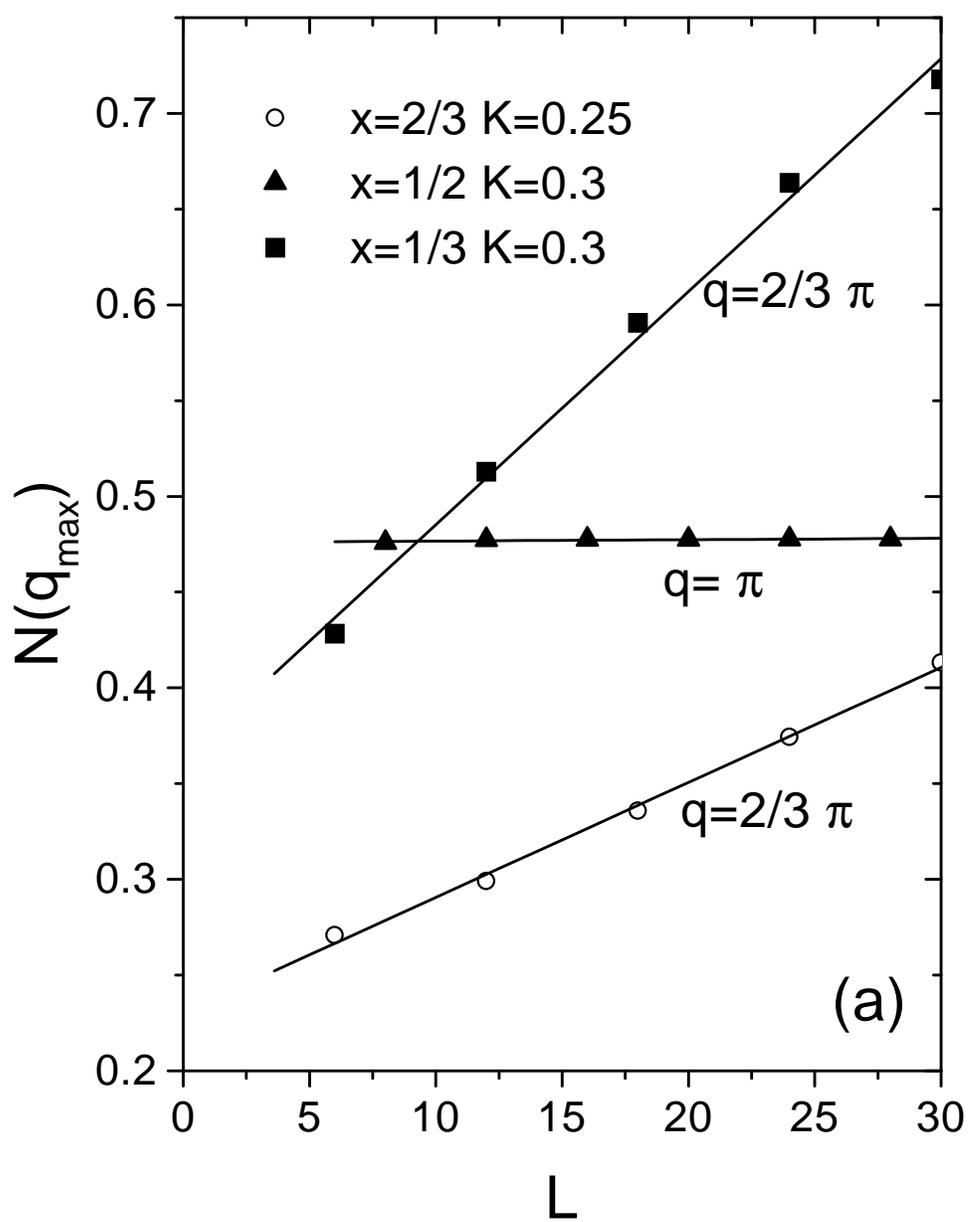
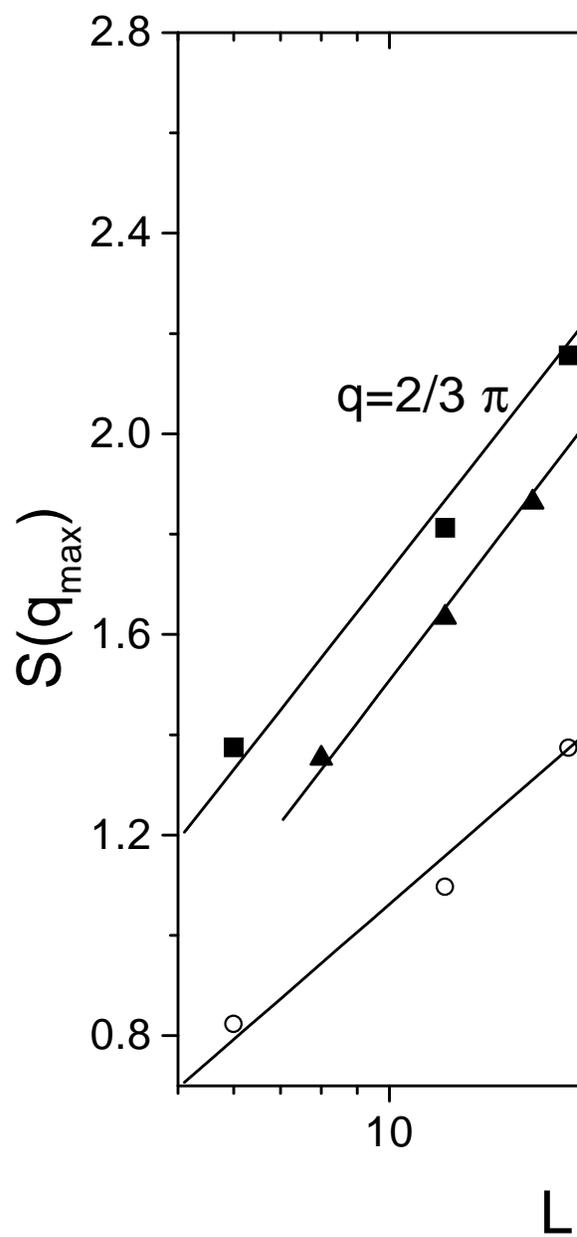
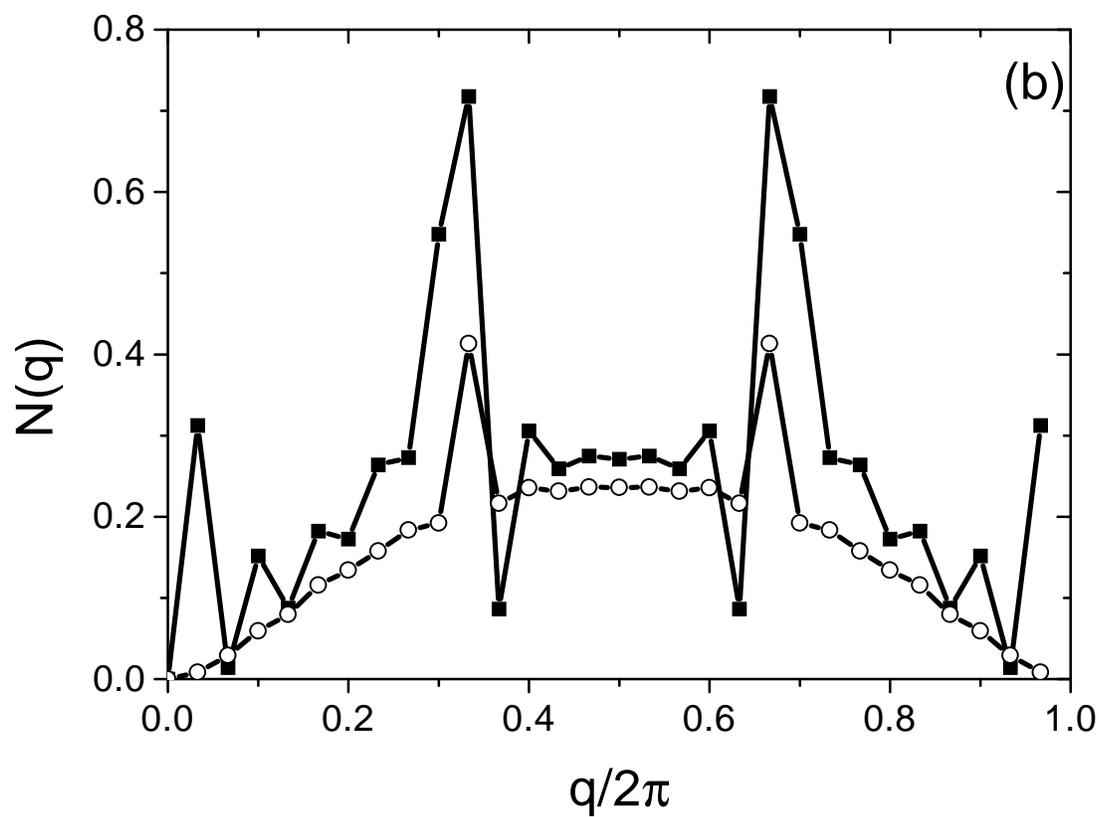
\end{document}